\documentclass[aps,twocolumn,prb,showpacs,superscriptaddress]{revtex4}
\usepackage{epsfig}
\usepackage{bm}
\def\be{\begin{equation}}
\def\ee{\end{equation}}
\def\bea{\begin{eqnarray}}
\def\eea{\end{eqnarray}}

\def\r#1{(\ref{#1})}
\def\nn{\nonumber\\}

\begin{document}
\title{Dynamical Structure Factor in Cu Benzoate and other
  spin-1/2 antiferromagnetic chains}
\author{Fabian H.L. Essler}
\affiliation{Department of Physics, Brookhaven National Laboratory,
 Upton, NY 11973-5000}
\author{A. Furusaki}
\affiliation{Yukawa Institute for Theoretical Physics, Kyoto University,
 Kyoto 606-8502, Japan}
\affiliation{Condensed-Matter Theory Laboratory, RIKEN,
 Wako, Saitama 351-0198, Japan}
\author{T. Hikihara}
\affiliation{Condensed-Matter Theory Laboratory, RIKEN,
 Wako, Saitama 351-0198, Japan}
\affiliation{Computational Materials Science Center,
 National Institute for Materials Science, Tsukuba, Ibaragi 305-0047,
 Japan}
\date{\today}

\begin{abstract}
Recent experiments of the quasi-one-dimensional spin-$\frac{1}{2}$
antiferromagnet Copper Benzoate established the existence of a
magnetic field induced gap. The observed neutron scattering intensity 
exhibits resolution limited peaks at both the antiferromagnetic wave
number and at incommensurate wave numbers related to the applied
magnetic field. We determine the ratio of spectral weights of these
peaks within the framework of a low-energy effective field theory
description of the problem.
\end{abstract}


\maketitle

\section{Introduction}
Recent experiments \cite{magn,dender} have investigated the behaviour of
the quasi-one-dimensional spin-1/2 antiferromagnet Copper Benzoate in
a magnetic field. Neutron scattering experiments
\cite{dender} established the existence of field-dependent
incommensurate low-energy modes in addition to low-energy modes at the
antiferromagnetic wave number. The incommensurability was found to
be consistent with the one predicted by the exact solution of the
Heisenberg model in a magnetic field. However, the system exhibited an
unexpected excitation gap, induced by the applied field. A theory for
this effect was put forward by Oshikawa and Affleck in
Ref.\ [\onlinecite{oa}]. Application of a uniform magnetic field
$\bm{H}$ induces a staggered field perpendicular to $\bm{H}$.
The staggered field is generated both by a staggered $g$-tensor
\cite{oshima} and a Dzyaloshinskii-Moriya (DM) interaction. 
The same physical mechanism has been found also in other materials.
\cite{other} The effective Hamiltonian describing such field-induced gap
systems is given by \cite{oa}
\be
\hat{H}=\sum_i J \bm{S}_i\cdot\bm{S}_{i+1}- H S_i^z + 
h (-1)^i S^x_i\ ,
\label{hamil}
\ee
where
\be
h=\gamma H\ .
\ee
The constant $\gamma$ is given in terms of the staggered
$g$-tensor \cite{oshima} and the DM
interaction. For Copper Benzoate the exchange constant is $J\approx
1.57\ {\rm meV}$ and the induced staggered field is much smaller than
the applied uniform field, $h\ll H$. 

As long as $h\ll J$, or equivalently as long as the field induced
gap $\Delta$ is much smaller than $J$, it is possible to describe the
low-energy degrees of freedom of \r{hamil} in terms of a massive,
relativistic quantum field theory. This low-energy effective theory is
obtained by abelian bosonization and is given by a Sine-Gordon model with
Hamiltonian density \cite{oa} 
\bea
{\cal
H}=\frac{v}{2}[(\partial_x\Phi)^2+(\partial_x\Theta)^2]
-\mu(h)\ \cos(\beta\Theta)\ .
\label{SGM}
\eea
Here $\Phi$ is a canonical Bose field, $\Theta$ is the dual field and
the coupling $\beta$ depends on the value of the applied uniform field
and has been calculated in Refs.\ [\onlinecite{et,fab,oa2}] by using 
the results of Ref. [\onlinecite{vladb}]. The spin
velocity $v$ also depends on $H$ and is shown in Fig.~9 of
Ref.\ [\onlinecite{oa2}]. It is useful to define
\be
\xi=\frac{\beta^2}{8\pi-\beta^2}.
\ee
The spectrum of the Sine-Gordon model \r{SGM} in the relevant range of
$\beta$ consists of a soliton-antisoliton doublet and several
soliton-antisoliton bound states called ``breathers.''\cite{oa,et}
The soliton gap as a function of $H$ and $h$ was determined in
Ref.\ [\onlinecite{oa2}] in the regime $\Delta\ll H$, where
\bea
&&\frac{\Delta}{J}\simeq\left(\frac{h}{J}\right)^{(1+\xi)/2}\nn
&&\qquad\times
\left[B \left(\frac{J}{H}\right)^{(2\pi-\beta^2)/4\pi}
\left(2-\frac{\beta^2}{\pi}\right)^{1/4}
\right]^{-(1+\xi)/2}
\label{deltaRG}
\eea
with $B=0.422169$.
Equation \r{deltaRG} is applicable as long as $H$ is sufficiently
smaller than $J$ or to be more precise as long as the magnetization is
small. For magnetic fields comparable to $J$ it is better to use the
following expression \cite{zam}
\bea
\frac{\Delta}{J}&\simeq&\frac{2\tilde{v}(H)}{\sqrt{\pi}}
\frac{\Gamma(\frac{\xi}{2})}
{\Gamma(\frac{1+\xi}{2})}
\left[\frac{c(H)\pi}{2\tilde{v}(H)}
\frac{\Gamma(\frac{1}{1+\xi})}{\Gamma(\frac{\xi}{1+\xi})}
\frac{h}{J}\right]^{(1+\xi)/2}\ .
\eea
Here  $\tilde{v}=v/(Ja_0)$ is the ``dimensionless spin velocity'',
$a_0$ is the lattice constant and $c(H)$ is given below.
The breather gaps are given by
\be
\Delta_n=2\Delta\sin\left(\frac{\pi\xi n}{2}\right),\quad
 n=1,\ldots,
\left[\frac{1}{\xi}\right]. 
\ee

\section{Dynamical Structure Factor}

The staggered/oscillating components of the spin operators are
expressed in terms of the continuum fields $\Phi$ and $\Theta$ as
\bea
S^z_n&\sim&(-1)^n\ a(H)\
\sin\left(\frac{2\pi}{\beta}\Phi-\frac{2\delta}{a_0} x\right)\ ,\nn
S^x_n&\sim&(-1)^n\ c(H)\ \cos(\beta\ \Theta)\ ,\nn
S^y_n&\sim&(-1)^n\ c(H)\ \sin(\beta\ \Theta)\ .
\label{ac}
\eea
Here $x=na_0$ and the incommensuration $\delta$ is determined from the
exact solution of the Heisenberg model in a uniform magnetic field
\cite{vladb,fab} [that is the Hamiltonian \r{hamil} for $h=0$]. The
amplitudes $a(H)$ and $c(H)$ are at present not known analytically,
but can be determined numerically with high accuracy. We note that
these amplitudes are also calculated in the absence of a staggered
field, the expectation being that the changes due to a small $h\ll H$
will be negligible. The data used in this work are obtained in the 
scheme of Refs.\ [\onlinecite{FH,HF2}]: We calculate the spin
polarization  $\langle S^z_n \rangle$ and the two-spin correlation
function $\langle S^x_n S^x_{n'} \rangle$ in the Heisenberg chain of
$200$ spins using the density-matrix renormalization group method, 
and then, fit them to analytic formulas obtained from the abelian 
bosonization taking $a(H)$ and $c(H)$ as fitting parameters.
The results as well as other parameters, which are determined 
exactly,\cite{et,oa2} are listed in Table \ref{tab:Amp} 
for several typical values of the magnetization $m$.

\begin{table}
\caption{
\label{tab:Amp}
Amplitudes $a$ and $c$, the spin velocity $v$, the coupling $\beta$, 
and the field $H$ as functions of the magnetization $m$.
The amplitudes are determined numerically 
except for $m = 0.5$ where exact values are shown.
The figures in parenthesis for $a$ and $c$ indicate the error 
on the last quoted digits.
}
\begin{ruledtabular}
\begin{tabular}{cccccc}
$m$ & $a$ & $c$ & $v$ & $\beta$ & $H$ \\
\hline
  0.02 &  0.591(3)  &   0.4937(3) &  1.54271 &  2.35016 &  0.17599 \\
  0.04 &  0.550(5)  &   0.4883(2) &  1.51707 &  2.31088 &  0.34214 \\
  0.06 &  0.520(4)  &   0.4863(2) &  1.48415 &  2.27738 &  0.50013 \\
  0.08 &  0.4947(6) &   0.4853(2) &  1.44425 &  2.24653 &  0.65001 \\
  0.10 &  0.475(1)  &   0.4847(2) &  1.39796 &  2.21731 &  0.79164 \\
  0.12 &  0.454(1)  &   0.4842(2) &  1.34593 &  2.18927 &  0.92489 \\
  0.14 &  0.437(2)  &   0.4835(2) &  1.28879 &  2.16216 &  1.04965 \\
  0.16 &  0.422(2)  &   0.4825(2) &  1.22720 &  2.13587 &  1.16589 \\
  0.18 &  0.4070(7) &   0.4810(2) &  1.16178 &  2.11029 &  1.27360 \\
  0.20 &  0.3938(8) &   0.4790(2) &  1.09314 &  2.08538 &  1.37287 \\
  0.22 &  0.3813(6) &   0.4764(2) &  1.02184 &  2.06107 &  1.46380 \\
  0.24 &  0.3700(8) &   0.4731(2) &  0.94844 &  2.03735 &  1.54656 \\
  0.26 &  0.3596(7) &   0.4690(2) &  0.87347 &  2.01418 &  1.62134 \\
  0.28 &  0.3499(4) &   0.4639(2) &  0.79741 &  1.99153 &  1.68839 \\
  0.30 &  0.3406(4) &   0.4578(2) &  0.72074 &  1.96940 &  1.74794 \\
  0.32 &  0.3330(2) &   0.4504(2) &  0.64387 &  1.94775 &  1.80030 \\
  0.34 &  0.3262(2) &   0.4416(2) &  0.56722 &  1.92658 &  1.84575 \\
  0.36 &  0.3200(3) &   0.4310(2) &  0.49116 &  1.90586 &  1.88462 \\
  0.38 &  0.3145(4) &   0.4183(2) &  0.41602 &  1.88559 &  1.91723 \\
  0.40 &  0.3094(2) &   0.4029(1) &  0.34212 &  1.86574 &  1.94390 \\
  0.42 &  0.3070(8) &   0.3841(1) &  0.26973 &  1.84631 &  1.96497 \\
  0.44 &  0.3058(2) &   0.3601(1) &  0.19912 &  1.82727 &  1.98079 \\
  0.46 &  0.3062(6) &   0.3284(1) &  0.13049 &  1.80863 &  1.99168 \\
  0.48 &  0.309(1)  &   0.2802(1) &  0.06407 &  1.79036 &  1.99797 \\
  0.50 &  0.3183    &   0         &  0       &  1.77245 &  2       \\
\end{tabular}
\end{ruledtabular}
\end{table}

\begin{figure}
\begin{center}
\noindent
\epsfxsize=0.4\textwidth
\epsfbox{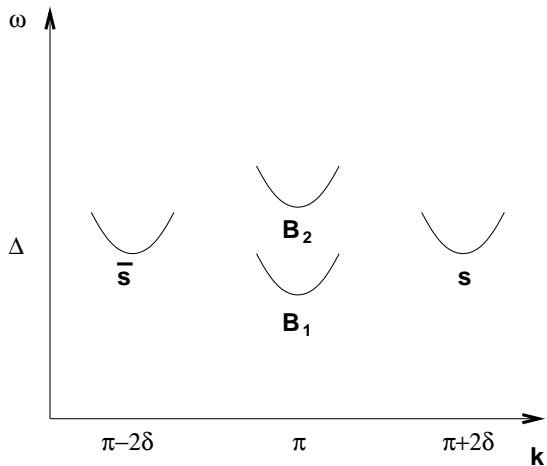}
\end{center}
\caption{\label{fig:modes}%
Schematic structure of the lowest-energy excited states relevant to
neutron scattering experiments. Soliton and antisoliton occur in the
vicinity of the incommensurate wave numbers $\pi\pm 2\delta$ and are
seen in $S^{zz}$,  whereas the first breather $B_1$ occurs in the
vicinity of $\pi$ and contributes to $S^{yy}$. At higher energies
further breather bound states as well as multi-particle scattering
continua are present.}  
\end{figure}

The inelastic neutron scattering intensity is proportional to
\be
I(\omega,{\bm k})\propto \sum_{\alpha,\beta}
\left(1-\frac{k_\alpha k_\beta}{{\bm k}^2}
\right)\ S^{\alpha\beta}(\omega,{\bm k})\ ,
\label{intensity}
\ee
where $\alpha, \beta = x,y,z$ and the dynamical structure factor
$S^{\alpha\beta}$ is defined by
\begin{eqnarray}
&&
S^{\alpha \beta} (\omega, k) = \sum_{l= 1}^N
\int_{-\infty}^{\infty} 
\frac{dt}{2\pi}e^{-i kla_0 + i \omega t}
\langle S^{\alpha}_{l+1} (t) S^{\beta}_{1} (0)
\rangle.
\nn
&&
\end{eqnarray}
Here $k$ denotes the component of $\bm{k}$ along the chain direction. A
schematic representation of which excited states will contribute to
the various components of the dynamical structure factor at
$k={\pi}/{a_0}$ and $k=(\pi\pm 2\delta)/{a_0}$ is shown in
Fig.~\ref{fig:modes}. At the antiferromagnetic wave number there are
several breather excitations and at higher energies multiparticle
continua.\cite{et} These contribute to the $xx$ and $yy$ components
of the dynamical structure factor, which have been determined in detail
in Ref.\ [\onlinecite{et}]. At the incommensurate wave numbers
$k=(\pi\pm 2\delta)/{a_0}$ there are soliton and antisoliton states
and at higher energies multiparticle scattering continua. In this paper
we calculate the single particle soliton/antisoliton contributions to
$S^{zz}$ and compare them to the dominant feature in the dynamical
structure factor, the contribution of the lightest breather bound
state $B_1$ to $S^{yy}$. 

\begin{figure}[ht]
\begin{center}
\epsfxsize=0.4\textwidth
\epsfbox{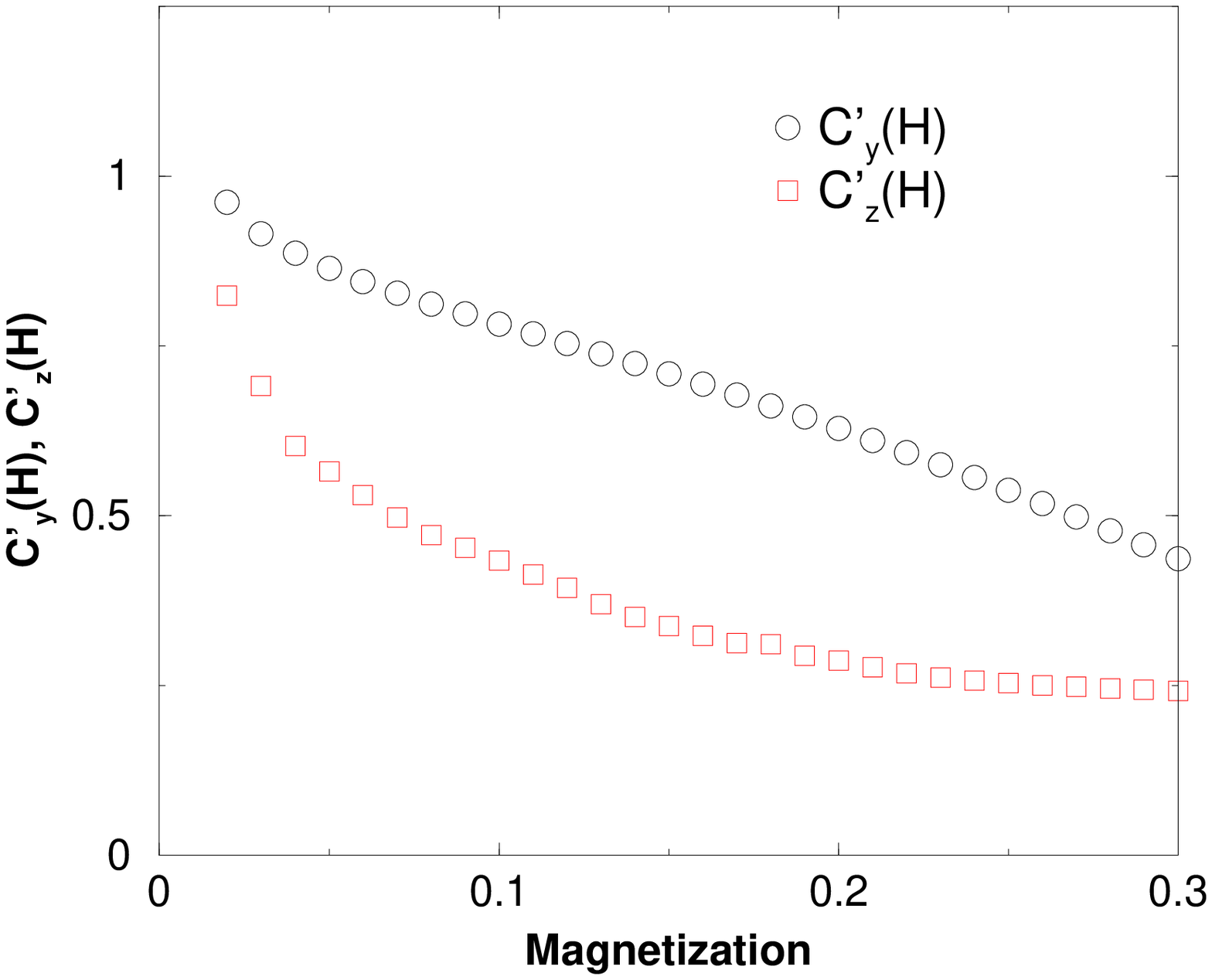}
\end{center}
\caption{\label{fig:chat}
Amplitudes $C'_{y,z}(H)$ as functions of the magnetization
.}
\end{figure}

The lightest breather $B_1$ has a gap $\Delta_1$ and contributes to
$S^{yy}$ as\cite{et}
\be
S^{yy}\left(\omega,\frac{\pi}{a_0}+q\right)\bigg|_{B_1}
=C_y(H)\ \delta\biglb(\omega^2-(vq)^2-\Delta_1^2\bigrb)\ ,
\ee
where
\begin{widetext}
\bea
C_y(H)&=&2\tilde{v}J\ c^2(H) \left[
2\cos(\pi\xi/2)\sqrt{2\sin(\pi\xi/2)}
\exp\left(-\int_0^{\pi\xi}\frac{dt}{2\pi}
\frac{t}{\sin t}\right)\right]^2
\left(\frac{\Delta}{J\tilde{v}}\frac{\sqrt{\pi}}{2}
\frac{\Gamma\biglb((1+\xi)/2\bigrb)}{\Gamma(\xi/2)}
\right)^{\beta^2/2\pi}\nn
&&\times\exp\left[
2\int_0^\infty\frac{dt}{t}\left(\frac{\sinh^2(2\beta^2t)}
{2\sinh(\beta^2 t)\sinh(8\pi t)\cosh[(8\pi-\beta^2)t]}
 -\frac{\beta^2}{4\pi}e^{-16\pi t}
\right)\right].
\eea
Here we have used the normalizations of Ref.\ [\onlinecite{lukyanov1}].
The leading contributions to the longitudinal structure factor at the
incommensurate wave numbers $k=(\pi\pm 2\delta)/{a_0}$ are due to
soliton and antisoliton. Using the results of
Ref.\ [\onlinecite{lukyanov2}] we obtain 
\bea
S^{zz}\left(\omega,\frac{\pi\pm 2 \delta}{a_0}+q\right)
\bigg|_{s,\bar{s}}&=&C_z(H)\
\delta\biglb(\omega^2-(vq)^2-\Delta^2\bigrb),\nn
\eea
where
\bea
C_z(H)&=&\frac{\tilde{v}J}{2}\ a^2(H)
\left(\frac{{\cal C}_1^4\xi}{4{\cal C}_2}\right)^{-1/4}
\left(\frac{\sqrt{\pi}\ \Delta\Gamma\left(\frac{3}{2}+\frac{\xi}{2}\right)}
{J\tilde{v}\Gamma\left(\frac{\xi}{2}\right)}\right)^{2\pi/\beta^2}\nn
&&\times\exp\biggl[
\int_0^\infty\frac{dt}{t}\biggl(\frac{\exp[-(1+\xi)t]-1}
{2\sinh(\xi t)\sinh[(1+\xi)t]\cosh(t)}
 +\frac{1}{2\sinh(t\xi)}-\frac{2\pi\ e^{-2t}}{\beta^2}
\biggr)\biggr].
\eea
\end{widetext}
Here the constants ${\cal C}_{1,2}$ are given by
\bea
{\cal C}_1&=&\exp\left(-\int_0^\infty \frac{dt}{t}
\frac{\sinh^2(t/2)\ \sinh[t(\xi-1)]}{\sinh(2t)\ \sinh(\xi t)
\ \cosh(t)}\right),\nn
{\cal C}_2&=&\exp\left(4\int_0^\infty \frac{dt}{t}
\frac{\sinh^2(t/2)\ \sinh[t(\xi-1)]}{\sinh(2t)\ \sinh(\xi t)}\right).
\eea
As was pointed out in Ref.\ [\onlinecite{oa2}], at $H=0$ the low-energy
effective theory of \r{hamil} is SU(2) symmetric. In our notations
this implies that
\bea
\lim_{H\to 0}\frac{C_y}{c^2(H)} = 2\lim_{H\to 0}\frac{C_z}{a^2(H)}.
\label{su2}
\eea
Equation \r{su2} is easily verified numerically. In order to evaluate
$C_{y,z}$ we need to know the constant of proportionality
$\gamma$ that relates the induced staggered field $h$ with the applied
uniform field $H$. This constant differs from compound to compound. On
the other hand, $\gamma$ enters the expressions for $C_{y,z}$
only via the soliton gap $\Delta$. Hence it is useful to isolate the
$\gamma$ dependence and consider the quantities
\bea
C'_y(H)&=&C_y(H)\ 
\left(\frac{\Delta}{J}\right)^{-\frac{\beta^2}{2\pi}}\ J^{-1}\ ,\nn
C'_z(H)&=&C_z(H)\
\left(\frac{\Delta}{J}\right)^{-\frac{2\pi}{\beta^2}}\ J^{-1}\ .
\eea
The amplitudes $C'_{y,z}(H)$ are shown as functions of the
magnetization in Fig.~\ref{fig:chat}.
\section{Copper Benzoate}

We are now in a position to determine the ratio between the spectral
weights of the first breather (seen in the transverse structure factor
$S^{yy}$) and the soliton (seen in the longitudinal structure factor
$S^{zz}$). In an ideal situation one would carry out measurements with
momentum transfers only along the $y$ and $z$ directions
respectively. In practice, the experiments on Copper Benzoate were
carried out with momentum transfers $\bm{k}_{1,2}$ respectively, where
\bea
({\bm k}_1\cdot{\bm a},{\bm k}_1\cdot{\bm b},{\bm k}_1\cdot{\bm c})&=&
2\pi\ (-0.3,0,1)\ ,\nn 
({\bm k}_2\cdot{\bm a},{\bm k}_2\cdot{\bm b},{\bm k}_2\cdot{\bm c})&=&
2\pi\ (-0.3,0,1.12)\ , 
\eea
Here ${\bm c}$ points along the chain direction and the
antiferromagnetic wave number corresponds to $2\pi/c$ ($c=6.30A$) as
there are two copper atoms per unit cell along the $c$-axis. To make
contact with our previous notations we need to set $a_0=c/2$.
The measurements with momentum transfers ${\bm k}_1$ and ${\bm k}_2$
probe the dynamical structure factor around $\pi/a_0$ and the
incommensurate wave number $(\pi+2\delta)/a_0$ respectively. In order
to make direct comparisons with the experiments we need to relate
the $(a,b,c)$ coordinate system describing the crystal axes to the
$(x,y,z)$ spin coordinates. By definition $z$ and $x$ are the
directions of the uniform and staggered fields respectively. In the
experiments of Ref.\ [\onlinecite{dender}] the uniform field was applied
along the $b$-direction. Based on a polarization analysis it was
suggested in Ref.\ [\onlinecite{oa2}] that the staggered field lies in the
$ac$ plane and encloses an angle of $\alpha=-72^\circ$ with the
$a$-axis. In the vicinity of $\pi/a_0$ the dominant contribution to
the structure factor comes from the transverse correlators. This
implies that
\bea
I(\omega,{\bm k}_1)&\propto& (0.083 \cos^2\alpha+0.917
\sin^2\alpha)S^{yy}\left(\omega,\frac{\pi}{a_0}\right) \nn
&&+ (0.083 \sin^2\alpha+0.917
\cos^2\alpha)S^{xx}\left(\omega,\frac{\pi}{a_0}\right)\nn
&\approx&0.84\ S^{yy}\left(\omega,\frac{\pi}{a_0}\right)
+0.16\ S^{xx}\left(\omega,\frac{\pi}{a_0}\right).
\label{I}
\eea
On the other hand at momentum transfer ${\bm k}_2$ the dominant
contribution to the structure factor is due to the longitudinal
component 
\bea
I(\omega,{\bm k}_2)&\propto&
 S^{zz}\left(\omega,\frac{\pi+2\delta}{a_0}\right)\ .
\label{I2}
\eea
As was pointed out in Ref.\ [\onlinecite{oa2}] there are unresolved issues
concerning the polarization analysis and the estimate of the angle
$\alpha$ should be regarded with some caution. It is possible to infer
$\alpha$ by analyzing other experiments such as specific heat and ESR
measurements. The analysis of the specific heat data suggests that
$\alpha\approx -82^\circ$,\cite{fab} which leads to a contribution of
about $90\%$ of $S^{yy}$ in \r{I}. 

The neutron scattering experiments of Ref.\ [\onlinecite{dender}] were
performed in a uniform magnetic field of $7\,$T, which corresponds to a
magnetization per site of $m\approx 0.06$. The breather and soliton
gaps were observed at
\be
\Delta\approx 0.22\ {\rm meV}\ ,\quad
\Delta_1\approx 0.17\ {\rm meV}.
\ee
Using the expression \r{deltaRG} for the soliton gap we can infer
the coefficient of proportionality between the uniform field $H$ and
the staggered field $h$ as $\gamma\approx 0.06$.
Taking $\gamma=0.06$ and $m=0.06$, Eq.\ \r{deltaRG} gives 
$\Delta= 0.215\ {\rm meV}$, $\Delta_1=0.171\ {\rm meV}$ and we will
use this set of parameters for our further analysis.

\begin{figure}[ht]
\begin{center}
\epsfxsize=0.4\textwidth
\epsfbox{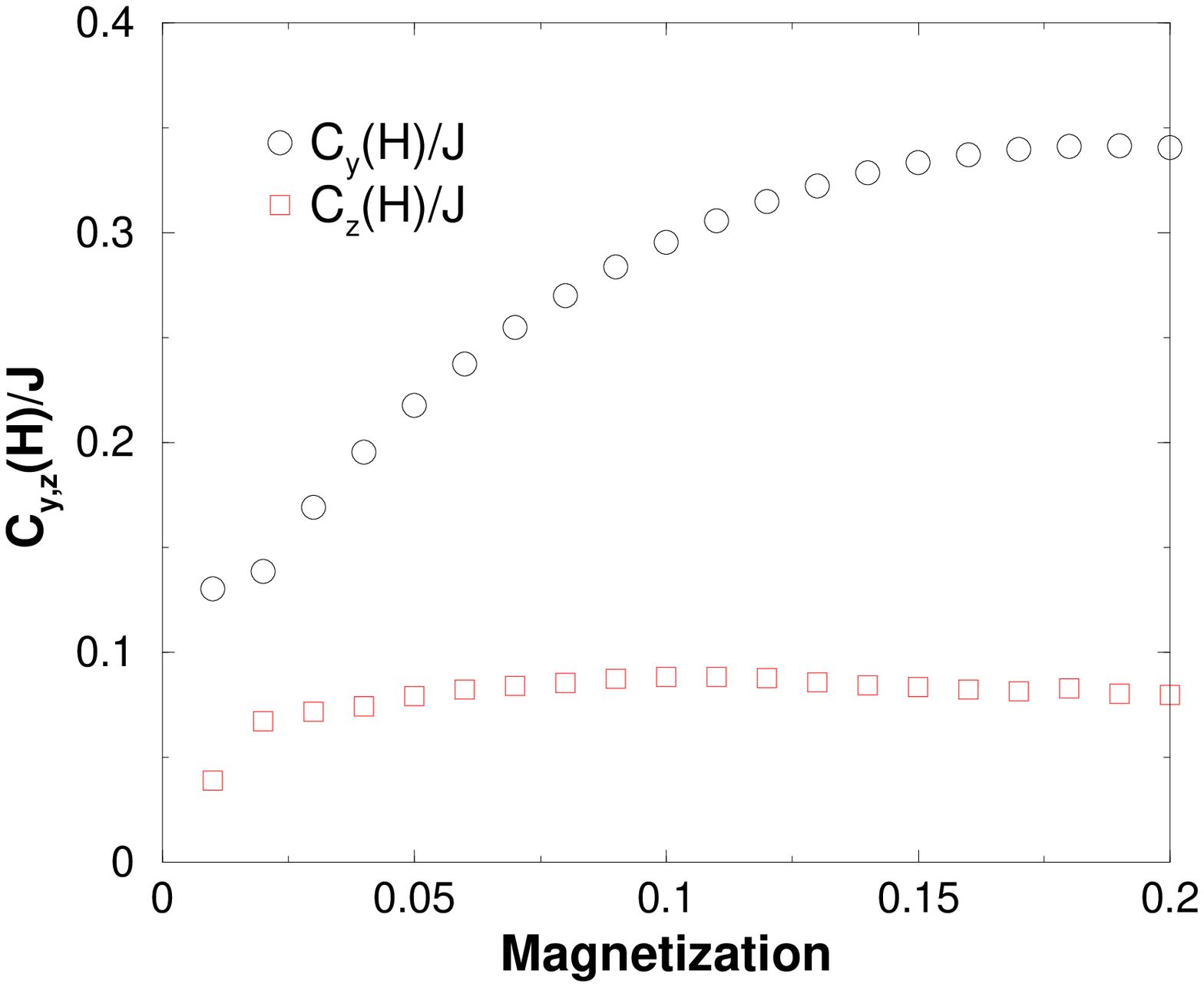}
\end{center}
\caption{\label{fig:ampl}
Amplitudes $C_{y,z}(H)$ as functions of the magnetization.}
\end{figure}

Under the above assumptions we may now determine the spectral weights
of the coherent soliton and breather peaks in the dynamical structure
factor. The results are shown in Fig.~\ref{fig:ampl}.

At a magnetization of $m=0.06$ we have
\be
\frac{C_y(H)}{C_z(H)}\approx 2.88. 
\ee

The spectral weights are obtained by integrating the respective
structure factors over frequency at fixed momentum, i.e.,
\bea
I_{B_1}&=&\int d\omega\ S^{yy}\left(\omega,\frac{\pi}{a_0}\right)
 \Biggl|_{B_1}\ ,\nn
I_{s}&=&\int d\omega\ S^{zz}\left(\omega,\frac{\pi+2\delta}{a_0}\right)
 \Biggl|_{s}\ .
\eea
The ratio of spectral weights between the first breather $I_{B_1}$ and
the soliton $I_s$ is approximately  
\be
\frac{I_{B_1}}{I_s}=\frac{C_y(H)}{C_z(H)}\frac{\Delta}{\Delta_1}
\approx 3.64.
\ee

In order to compare to experiment, we need to take into
account the different momentum transfers in the measurements of
$S^{yy}$ and $S^{zz}$ respectively. From \r{I} 
and \r{I2} we arrive at the following theoretical prediction for the
ratio of intensities 
\bea
R=0.84 \frac{I_{B_1}}{I_{s}}\approx 3.06.
\eea
The experimentally observed \cite{dender} ratio of peak heights
between the breather and soliton peaks is approximately $2.8$. This is
in reasonable agreement with our result. For a better comparison one
should take into account the resolution function of the instrument in
both momentum and energy, but this goes beyond the scope of our
present analysis.

\acknowledgments
Work at Brookhaven National Laboratory was carried out under contract
number DE-AC02-98 CH10886, Division of Material Science,
U.S. Department of Energy. 
The research of A.F. was supported in part by a Grant-in-Aid for
Scientific Research on Priority Areas from the Ministry of Education,
Culture, Sports, Science and Technology (Grant No. 12046238).


\begin{thebibliography}{99}

\bibitem{magn}
D.C. Dender, D. Davidovi\'c, D.H. Reich, C. Broholm, K. Lefmann and
G. Aeppli, Phys. Rev. B {\bf 53}, 2583 (1996).

\bibitem{dender}
D.C. Dender, P.R. Hammar, D.H. Reich, C. Broholm and G. Aeppli,
Phys. Rev. Lett. {\bf 79}, 1750 (1997).

\bibitem{oa}
M. Oshikawa and I. Affleck, Phys. Rev. Lett. {\bf 78}, 1984 (1997).

\bibitem{oshima}
K. Oshima, K. Okuda and M. Date, J. Phys. Soc. Jpn. {\bf 41}, 475
(1976), J. Phys. Soc. Jpn. {\bf 44}, 757 (1978).

\bibitem{other}
R. Feyerherm \textit{et al}., J. Phys.: Condens. Matter {\bf 12},
 8495 (2000);
M. Oshikawa \textit{et al.}, J. Phys. Soc. Jpn. {\bf 68}, 3181 (1999).

\bibitem{oa2}
I. Affleck and M. Oshikawa, Phys. Rev. B {\bf 60}, 1038 (1999).

\bibitem{et}
F.H.L. Essler and A.M. Tsvelik, Phys. Rev. B {\bf 57}, 10592 (1998).

\bibitem{fab}
F.H.L. Essler, Phys. Rev. B {\bf 59}, 14376 (1999).

\bibitem{vladb}
V.E. Korepin, A.G. Izergin, and N.M. Bogoliubov, {\em {Quantum Inverse
  Scattering Method, Correlation Functions and Algebraic Bethe Ansatz}}
  (Cambridge University Press, Cambridge UK, 1993).

\bibitem{zam}
Al.B. Zamolodchikov, Int. Jour. Mod. Phys.{\bf A10}, 1125 (1995).

\bibitem{FH}
T. Hikihara and A. Furusaki, Phys. Rev. B {\bf 63}, 134438 (2001).

\bibitem{HF2}
T. Hikihara and A. Furusaki, unpublished.

\bibitem{lukyanov1}
S. Lukyanov, Mod. Phys. Lett. {\bf A12}, 2911 (1997).

\bibitem{lukyanov2}
S. Lukyanov and A. Zamolodchikov, Nucl. Phys. {\bf B607}, 437 (2001).


\end{thebibliography}
\end{document}